# Local characterization of superconductivity in BaFe$_2$(As$_{1-x}$P$_x$)$_2$

By

Y. Lamhot,[1] A. Yagil,[1] N. Shapira,[1] S. Kasahara,[2] T. Watashige,[2] T. Shibauchi,[3] Y. Matsuda[2] and O. M. Auslaender[1*]

[1]Department of Physics, Technion – Israel Institute of Technology, Haifa 32000, Israel

[2]Department of Physics, Kyoto University, Kyoto 606-8502, Japan

[3]Department of Advanced Materials Science, University of Tokyo 5-1-5 Kashiwanoha, Kashiwa, Chiba 277-8561, Japan

[*]Address correspondence to ophir@physics.technion.ac.il

PACS numbers: 07.79.Pk, 74.70.Xa, 74.25.Dw, 74.40.Kb

## ABSTRACT

We use magnetic force microscopy (MFM) to characterize superconductivity across the superconducting dome in BaFe$_2$(As$_{1-x}$P$_x$)$_2$, a pnictide with a peak in the penetration depth ($\lambda_{ab}$) at optimal doping (x$_{opt}$), as shown in sample-wide measurements. Our local measurements show a peak at x$_{opt}$ and a $T_C$ vs. $\lambda_{ab}^{-2}$ dependence similar on both sides of x$_{opt}$. Near the underdoped edge of the dome $\lambda_{ab}$ increases sharply suggesting that superconductivity competes with another phase. Indeed MFM vortex imaging shows correlated defects parallel to twin boundaries only in underdoped samples and not for x ≥ x$_{opt}$.

The origin of superconductivity in the iron-based materials is still under debate although there is mounting evidence for the role of magnetic order and fluctuations [1–4]. For instance, it is well established that the parent compound for the pnictides is a metal with spin-density-wave (SDW) order and that doping by electrons, holes or isovalently gives rise to superconductivity and suppresses the magnetic order and an associated structural phase transition [5–9]. Moreover, the optimal doping for the superconducting transition temperature ($T_C$) is only slightly higher than the maximum doping for which SDW order has been observed. This implies that there is a range of doping for which magnetic order coexists with superconductivity [4,10,11], which can give rise to unconventional superconductivity. When the SDW transition temperature $T_N \to 0\,\text{K}$ a quantum critical point (QCP) may be present [12] at the doping at which superconductivity changes its nature [13–18]. Several recent experiments on the isovalently doped pnictide BaFe$_2$(As$_{1-x}$P$_x$)$_2$ show evidence consistent with this [18], including mass enhancement [19] and the behavior of transport coefficients [17,20]. Particularly striking is a recent observation in sample-wide measurements of a peak in the penetration depth at $T \ll T_C$ near optimal doping [21]. As the absolute value of the penetration depth is a direct measure of both the superconducting carrier density ($n_S$) and the effective mass [22], such a peak may hint at a QCP within the superconducting dome [23–25].

Here we report magnetic force microscopy (MFM) measurements of the local absolute value of the in-plane penetration depth ($\lambda_{ab}$) in BaFe$_2$(As$_{1-x}$P$_x$)$_2$. At the location where we measure $\lambda_{ab}$ we also measure the local $T_C$ in order to determine the relationship between these two fundamental superconducting properties. In addition we use MFM to map the location of superconducting vortices, which can become trapped by defects in the material. This allows us to learn about correlated defects that may arise as a result of structural and magnetic phase transitions.

Our samples were high-quality single crystals grown by the self-flux method and annealed in vacuum. Before each cool-down we cleaved each sample and chose pieces with a large, flat face suitable for scanning. After the measurement we analyzed each sample by EDS (Energy-Dispersive x-ray Spectroscopy) to determine x at the actual scanned surface at several different locations using a measurement area of $\approx 50 \times 50\ \mu\text{m}^2$ [26]. In addition to x, the EDS reported in all cases the expected atomic compositions for Ba (19.0-21.0%) and Fe (38.4-41.0 %). The EDS results also indicate that the surfaces we scanned were devoid of contaminants. This also implies an absence of a dead-layer at the surface.

Our measurements were done by low temperature magnetic force microscopy (MFM) on samples covering a range of doping that spans the entire superconducting dome: from underdoped (x=0.22, 0.26, 0.29), through optimal doping (x=0.30) to overdoped (x=0.33, 0.46, 0.55), as listed in the table in [27]. For $x \approx 0.29$ we looked at two different pieces from the same crystal whereas for x=0.22 we measured samples from two distinct growth batches (see table in [27]). The scatter of the values we obtained for x by EDS gives a variance of ≤ 1% for all samples except for both the x=0.22 samples which had a variance of 2%. Because our signal is affected by a region in the sample only up to a few micrometers in diameter and only a few hundred nanometers deep, on the order of $\lambda_{ab}$, our results are less sensitive to inhomogeneity than measurements which average over the whole sample [28,29]. The locality also allows us to check homogeneity by comparing measurements from different areas in each sample.

In our setup the magnetic MFM tip [30] is subjected to forces due to the Meissner response from the superconducting sample, the magnetic field from vortices and magnetic fields from other sources, if they are present. We minimize the electrostatic forces between the tip and the sample by compensating for the contact potential difference. We work with frequency modulated MFM in which the forces on the tip shift the resonant frequency of the cantilever holding it: $\Delta f = C - (f_0/2k_0)(\partial F_z/\partial z)$ ($f_0$ is the cantilever's natural resonance frequency, $k_0$ is the spring constant, $z$ is the direction normal to the sample surface and $C$ is a constant offset) [31].

For $\lambda_{ab}$ measurements we cool the sample in low magnetic field and find an area without vortices and visible defects. For this we scan the sample by moving the tip in a raster pattern parallel to the surface while recording $\Delta f(x, y)$ at a constant height $z$. After finding a suitable area we perform a 'touchdown' – we bring the tip to the surface and record $\Delta f$ vs. $z$. We performed all touchdown measurements at magnetic fields on the order of B<0.5 G with the nearest vortex at least 5 $\mu m$ away. Our calculations [27] show that the vortex contribution is much smaller than the Meissner response if the vortex-tip planar distance is greater than 0.5 $\mu$m. From such 'touchdown' curves we both estimate the local $T_C$ to an accuracy as good as $\pm 0.25$ K (see table in [27]) from the lowest temperature at which we do not see the Meissner effect (see Fig. 1) and determine $\lambda_{ab}$ by fitting to a model of the tip (see discussion in [27]). Both our model and our fitting procedure are refinements of previous work [28,29]. An essential part of our procedure to determine $\lambda_{ab}$ is a simple model for the magnetic tip. We assume a thin magnetic coating and a tip shape characterized by a small number of parameters (Fig. 2A). We set two of these -- $\theta$ (the cone half angle) and $h$ (length of the cone truncation) -- from SEM (Scanning Electron Microscopy) imaging of each tip (top inset, Fig. 2A). We determine the third parameter, $H$ (the effective magnetic height of the tip) by optimizing the fitting process as explained below. Using these parameters we fit for $m_0$ (which gives the magnetic strength of the tip), $\lambda_{ab}$ and the overall offset $C$. Because superconductivity in the underdoped samples is anisotropic the value we obtain for $\lambda_{ab}$ is the average between $\lambda_a$ (the decay length for currents flowing along the a-axis) and $\lambda_b$ (the decay length for currents flowing along the b-axis). For other samples $\lambda_{ab} = \lambda_a = \lambda_b$.

It is difficult to determine $H$ directly because it is influenced by the magnetic domain structure of the tip, which can vary from cool-down to cool-down [27]. The fit itself (Eq. S3 in [27]) gives different results for different values of $H$ and depends on the range of data we use for fitting, $z_{span}$. All of this can be seen in Fig. 2B, which also shows that there is a value for $H \equiv H^*$ for which the fit results have minimal sensitivity to $z_{span}$ ($H^* = 16$ $\mu$m in Fig. 2B, cf. arrow). The fit with $H = H^*$ gives the values of $\lambda_{ab}$ that we are reporting. We estimate the accuracy of the result by repeating the measurement several times at each point, from running the fit with parameters spanning the range of uncertainty in tip geometry and also from the error output of the fit routines [32].

Our results for the local values of $\lambda_{ab}$ and $T_C$ are shown in Fig. 3. The measurements for $\lambda_{ab}$ were done at the base temperature of our system ($T_{base} \approx 4.5 K$). The $T = 0$ K value can be extrapolated from these using previous measurements of the temperature dependence of $\lambda_{ab}$ [21] but the difference is less than 25 nm. With the exception of the most underdoped samples (x=0.22) both $\lambda_{ab}$ and $T_C$ are uniform across each sample. The result for $T_C$ is the usual dome-like dependence on x (Fig. 3A), although the curvature at optimal doping appears larger than previously reported for BaFe$_2$(As$_{1-x}$P$_x$)$_2$ [21].

For $\lambda_{ab}$ our local results clearly show a peak in the vicinity of optimal doping (x=0.3) as previously reported in sample-wide measurements [21]. In addition we see a sharp increase in $\lambda_{ab}$ at the lowest doping (x=0.22) that is reminiscent of both local and sample-wide measurements in the Ba(Fe$_{1-x}$Co$_x$)$_2$As$_2$ family [29,33]. Such an enhancement is expected in mean field theory far from the possible QCP near optimal doping [23,24]. Unlike the enhancement near optimal doping, which reflects an enhancement of the effective mass [19], the enhancement of $\lambda_{ab}$ at low doping is likely due to the suppression of $n_s$ as the SDW gap becomes larger and $T_C$ is suppressed with decreasing doping [34]. Thus the enhancement of $\lambda_{ab}$ appears to indicate the microscopic coexistence of SDW and superconductivity.

Our local measurement of $T_C$ and $\lambda_{ab}$ at the same location allows us to explore the relationship between them independent of x, as we show in Fig. 3B. With the exclusion of the x=0.22 data, we see the same dependence of $T_C$ on $\lambda_{ab}^{-2}$ on both overdoped and underdoped sides of optimal doping. This is highlighted by the dashed line Fig. 3B which shows a dependence of $T_C$ on $\lambda_{ab}^{-2}$ that is at complete odds with the Uemura linear relationship [35].

In the x=0.22 samples we find more variation from point to point in x, $\lambda_{ab}$ and $T_C$ as well as a strong enhancement of $\lambda_{ab}$. The large scatter is likely a consequence of the strong dependence of $\lambda_{ab}$ and $T_C$ on x and the variation of x from point to point. This can be seen in Fig. 3B, where we show that x=0.22 samples can have $T_C$ =10.5 K with $\lambda_{ab} \approx$ 700 nm and $T_C \approx$ 13 K with $\lambda_{ab} \approx$ 450 nm. Figure 3B highlights how different the x=0.22 results are from the results for the rest of the dopings. Unlike the single branch we see for x $\neq$ 0.22 the results for x=0.22 show a positive correlation between $T_C$ and $\lambda_{ab}^{-2}$. Possibly this behavior is another manifestation mean field effects governing the behavior far from the possible QCP near optimal doping.

While here we report a peak in $\lambda_{ab}$ at optimal doping that is similar to the peak reported in sample-wide measurements, there are significant differences between the results. First, our local measurements show a sharp increase of $\lambda_{ab}$ at low doping. Second we show a single valued dependence of $T_C$ on $\lambda_{ab}^{-2}$ where the sample-wide measurements showed two branches meeting at the lowest $\lambda_{ab}^{-2}$. Last, here we report somewhat larger values for $\lambda_{ab}$. While on the overdoped side the enhancement is compatible with our higher measurement temperature ($\approx 20$ nm), the enhancement in the vicinity of the peak ($\approx 60$ nm) is too large to be explained by temperature alone.

Finally we present images of arrays of superconducting vortices obtained at a field of a few Gauss. These can map defects that affect superconductivity locally by changing where vortices are pinned as a sample is cooled through $T_C$. We obtain such vortex decoration scans by rastering the magnetic MFM tip over the sample so that it interacts with the vortices. In such a scan vortices show up as oval, regular, features if the tip-vortex interaction is not strong enough to overcome the local pinning potential and as streaked features if it is [37]. In particular such a map can show the twin boundaries [6,8,36] that may accompany the structural phase transition from the high temperature tetragonal phase to the low temperature orthorhombic phase [5–7] that occurs in close proximity to $T_N$ [5,9]. Figure 4 shows results from vortex imaging (additional images are in [27]). Since we could not dislodge vortices at T<5 K in any sample with x>0.22 we conclude that for those samples vortex pinning was strong. In samples with x ≥ 0.28 we observed vortex arrays with no evidence for correlated defects such as twin boundaries, which should appear as straight lines [36]. The vortex configuration showed some degree of orientational order (cf. Fig. 4A) with rather uniform vortex-vortex spacing. This indicates that pinning was weak enough for vortex-vortex interactions to play a role in determining vortex positions. The vortex configuration was different in nature in the underdoped samples. In both x=0.26 samples we could clearly see lines of vortices (Fig. 4B). This indicates the presence of correlated defects (see e.g. [36,38]). It is possible that these are the twin boundaries one expects because at x=0.26 $T_C$ is below the structural phase transition (50 K at x=0.26 [5,9,20]). Indeed room temperature EBSD (Electron Back Scatter Diffraction) shows that the vortex lines are at 45° to the crystal axes at low temperature (black arrows in Fig. 4B), supporting the notion that we see vortices trapped on or near twin boundaries. At our lowest doping (x=0.22) we could not see vortices clearly (Fig. 4C) on either of the samples from two growth batches. The vortex images were worse near positions where the value of $\lambda_{ab}$ we extracted from touchdowns was relatively large (implying weaker tip-vortex interaction). In x=0.22 samples with vortices that were discernable they moved during the scan even at $T_{base}$. This indicates that there were no positions with dominant pinning. Even with the tip far from the surface to reduce the tip-vortex interaction vortices appeared blurry, another indication of vortex motion, which became more pronounced when the tip was closer to the surface. In none of the x=0.22 samples could we see evidence of correlated defects. This means that they were either absent, too dense for us to resolve or gave rise to pinning that was too weak for our tip.

In conclusion, we have used MFM for local measurements of $T_C$ and $\lambda_{ab}$ as well as for imaging vortices in single crystal BaFe$_2$(As$_{1-x}$P$_x$)$_2$ samples across the superconducting dome. The local $\lambda_{ab}$ dependence on doping has a peak at optimal doping that was previously only seen in sample-wide measurements [21]. We see a new feature in BaFe$_2$(As$_{1-x}$P$_x$)$_2$ - a divergence of $\lambda_{ab}$ at low doping that is likely a consequence of microscopic coexistence between SDW and superconductivity. The local $T_C$ measurements give the expected dome-shaped dependence on doping. Our results indicate that on both sides of optimal doping $T_C$ has the same $\lambda_{ab}^{-2}$ dependence. Vortex imaging showed an absence of correlated pinning in overdoped samples. In very underdoped samples (x=0.22) we found spread out vortices which were very weakly pinned. In mildly underdoped samples (x=0.26) we saw clear lines of vortices indicating correlated defects were present. Presumably these are twin-boundaries, further indicating that at this doping samples may be in a mixed SDW-superconductivity state.

## Acknowledgements

We would like to thank E. Berg, J. E. Hoffman, A. Kanigel, A. Keren, M. Khodas and D. Podolsky for discussions, A. Ribak and A. Brenner for help with EDS and EBSD. Y.L. acknowledges support from the Technion Russell Berrie Nanotechnology Institute (RBNI). This work was supported by the EU Seventh Framework Programme (FP7/2007-2013) under grant agreement no. 268294 and by the Israel Science Foundation (grant no. 1051/10).

**Figure 1**

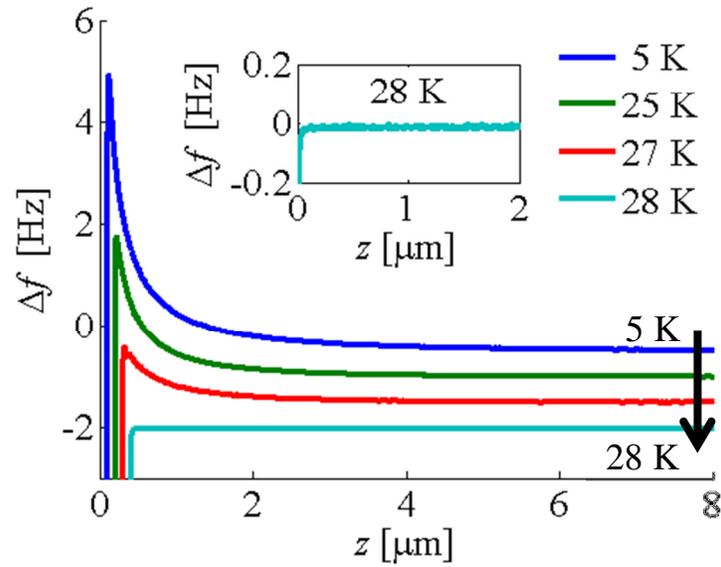

**Figure 1:** Touchdown curves for the x=0.29 sample at one position as a function of temperature. The curves are offset for clarity both horizontally (by 0.1 $\mu$m) and vertically (by 0.5 Hz). The left vertical in each curve indicates the surface of the sample (i.e. $z=0$ $\mu$m pre-offset). **Inset:** zoom-in on the $T$=28 K measurement (no offsets). The absence of a Meissner signal in this touchdown and its presence at 27 K (main panel, red curve) imply that $T_C = 27.5 \pm 0.5$ K.

**Figure 2**

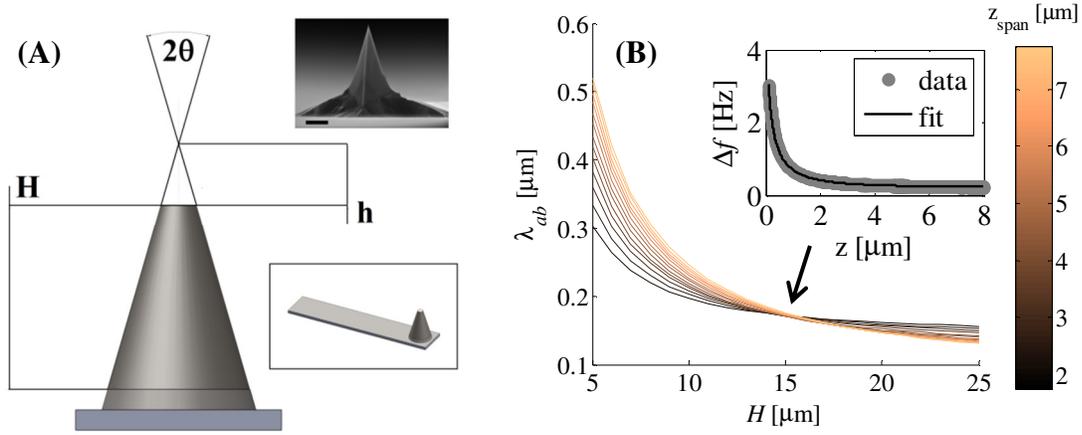

**Figure 2:** **(A)** Illustration of the model for the tip depicting $\theta$ (the cone half angle), $h$ (the length of the cone truncation, shown out of scale) and $H$ (the effective magnetic height of the tip). **Top Inset:** SEM image of a tip with a 4 $\mu$m scale bar. **Bottom Inset:** An illustration of the tip and the cantilever holding it. The cantilever is 220 $\mu$m long, 30 $\mu$m wide and approximately 3 $\mu$m thick. **(B)** The value of $\lambda_{ab}$ from fitting as a function of $z_{\text{span}}$ and $H$. The arrow points to the optimal $H = H^* = 16\,\mu$m for this case. In other cases we get comparable values as well as higher ones, up to many tens of microns [27]. **Insert**: The raw data and the fit at the optimal $H^* = 16\,\mu$m, which gives: $\lambda_{ab} = 170 \pm 5$ nm.

# Figure 3

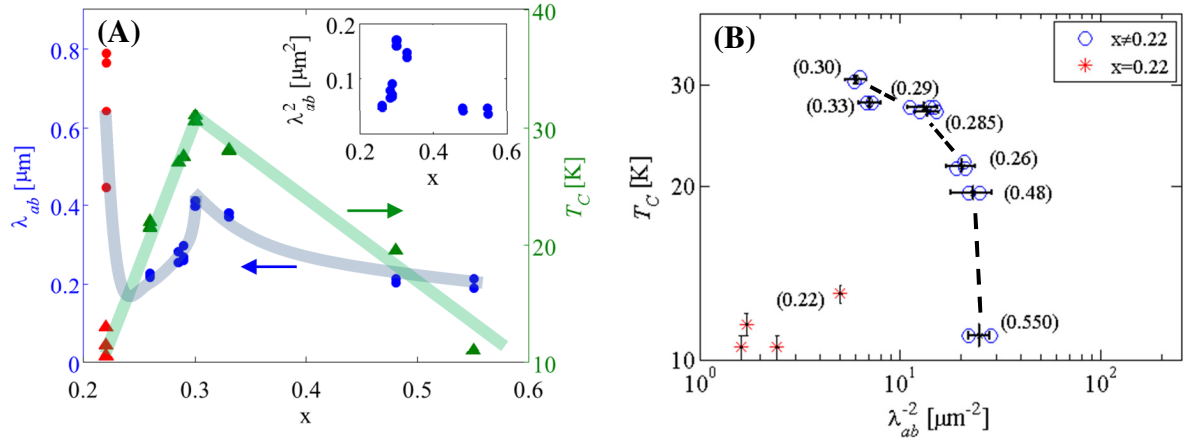

**Figure 3:** Results of the local measurement of $\lambda_{ab}$ and $T_C$ (all lines are guides to the eye). **(A)** $\lambda_{ab}$ (full circles, left axis, light blue line as guide to the eye) and $T_C$ (triangles, right axis, green line as guide to the eye) as a function of x. The data for x=0.22 is shown in red. The error bars are smaller than the symbols. **Inset:** $\lambda_{ab}^2$ as a function of x excluding $x=0.22$ showing the sharp peak at optimal doping. **(B)** $T_C$ as a function of $\lambda_{ab}^{-2}$ for all samples and from all touchdown locations. There are seven blue clusters of data points for $x>0.22$ samples, each marked by its value of x as determined from EDS (cf. table in [27]). In every cluster we show the mean values of $T_C$ and $\lambda_{ab}^{-2}$. The error bars take into account both the scatter due to repeat measurements as well as fit errors. The dashed line is a guide to the eye. Data for x=0.22 is shown in red asterisks with error bars for the value of $T_C$. The errors for $\lambda_{ab}^{-2}$ on each asterisk are smaller than the symbols.

# Figure 4

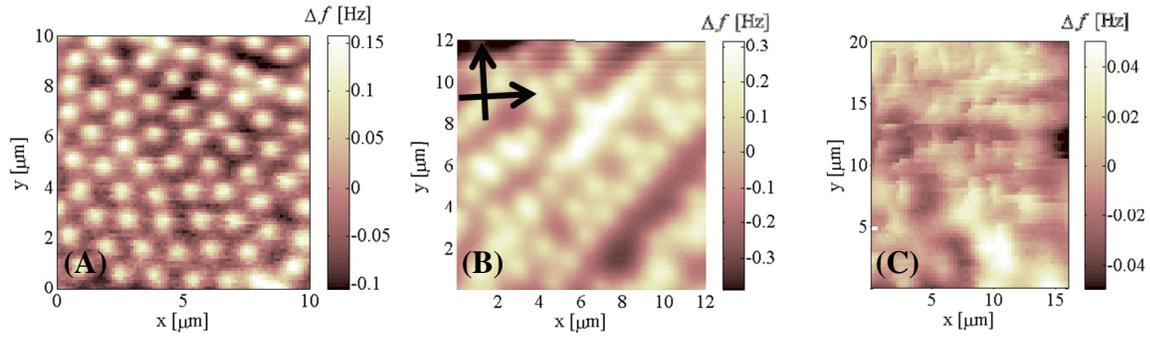

**Figure 4:** Vortex decoration scans. **(A)** x=0.33 sample with *z*=600 nm, T=4.45 K, B=9 G. **(B)** x=0.26 sample with *z*=650 nm, T=4.45 K, B=5 G. Crystal axes *ab* in this sample are shown in black arrows. **(C)** x=0.22 sample with *z*=850 nm, T=4.5 K, B=5 G.

# Supplementary information for

# Local characterization of superconductivity in BaFe$_2$(As$_{1-x}$P$_x$)$_2$

By

Y. Lamhot, A. Yagil, N. Shapira, S. Kasahara, T. Watashige, T. Shibauchi, Y. Matsuda and O. M. Auslaender

## Contents

## 1. Fitting touchdowns and a model for the MFM tip

As described in the main text, we fit touchdown data $(\Delta f(z))$ to a model of a magnetic tip approaching the surface of a superconductor. In this section we describe the model for the tip and the resulting theoretical touchdown curve.

The origin of our measured signal $\Delta f(z) \propto -\partial F_z/\partial z$ ($F_z$ is the $z$-component of the force acting on the tip) is the Meissner repulsion from the sample that the tip experiences. The truncated cone model (TCM), our model for the tip (Fig. 1 in the main text), assumes the magnetic coating of the tip is thin and is described by a small number of geometric parameters: $\theta$ (the cone half-angle), $h$ (the length of the truncated cone edge) and $H$ (the effective magnetic height of the tip) [1]. We also assume that the total magnetization is along the axis of the cone. Choosing coordinates in which the apex of the tip is at the origin the tip magnetization in the model is:

$$\vec{M}_{TCM}(x,y,z) = m_0 \, \delta\left(\sqrt{x^2+y^2} - R(z)\right) \Theta(z)\Theta(H-z)\hat{z}, \quad [S1]$$

where the magnetic strength of the tip is accounted for by $m_0$ and $R(z) = (z+h)\tan\theta$ (the delta function places magnetic dipoles on a thin shell for each value of $z$). The Heaviside step functions ($\Theta$) limit the magnetization to cover $0 \leq z \leq H$. As explained in the main text the MFM signal comes from the $z$-component of the gradient of the force in the $z$-direction, $\Delta f = -(f_0/2k)(\partial F_z/\partial z)$. We can calculate $\partial F_z/\partial z$ explicitly from the model using the Meissner repulsion that a single magnetic dipole experiences [2,3]:

$$\frac{\partial F_z}{\partial z} = -\frac{\mu_0}{2\pi} \int_0^\infty dk\, k^4\, e^{-2kz}\, G(k\lambda_{ab}) \int_{V_{tip}} d^3r'_{tip}\, M(\vec{r}'_{tip}) \int_{V_{tip}} d^3r''_{tip}\, M(\vec{r}''_{tip})\, e^{-k(z'_{tip}+z''_{tip})}\, J_0\left(\left|\vec{\rho}'_{tip} - \vec{\rho}''_{tip}\right|k\right). \quad [S2]$$

Here $\vec{\rho}_{tip} \equiv (x_{tip}, y_{tip})$ and $G(\xi) \equiv \dfrac{\sqrt{\xi^2+1}-\xi}{\sqrt{\xi^2+1}+\xi} = 1 + 2\xi^2 - \xi\sqrt{\xi^2+1}$. The structure of Eq. S2 is that of integrating over "two" tips: the so called original and "image" tips.

We find the force-gradient dependence on height $z$ by setting the magnetization $M(\vec{r}) = \vec{M}_{TCM}(x,y,z)$ from Eq. S1 in Eq. S2. It is possible to fit directly to Eq. S2 but with the Bessel function and the integration that would make each iteration in the fit slow. We wanted to accelerate the fitting process, especially since we used bootstrapping, which runs the fit many times. For that we needed a simple formula, which we can obtain with a few approximations. First, we replace $G(k\lambda_{ab})$ by its small $k\lambda_{ab}$ approximation, $\exp[-2k\lambda_{ab}]$. This approximation is valid when $z \gg \lambda_{ab}$ because the exponential term $\exp[-kz]$ in Eq. S2 gives an effective upper cutoff to the $k$-integration: only $k \ll z^{-1}$ contributes to the integral so $k\lambda_{ab} \ll 1$. The end result of this approximation is that when $z$ is large enough the force that we measure is due to the interaction between the original magnetic tip and an image created by mirroring the original tip at a plane a distance $\lambda_{ab}$ beneath the surface of the sample [2]. This image is of course formed by currents in the superconductor. Because we want to use the approximation $z \gg \lambda_{ab}$ we limit our analysis to data with $z > 1.2\lambda_{ab}$. In addition to simplifying the analysis, this restriction allows us to neglect other forces that become important when the tip is very close to the surface. The influence of such forces can be easily seen in Fig. 1 in the main text, where close to the surface the signal drops quickly to negative values due to van-der-Waals and remnant electrostatic interaction between the tip and the sample.

The second approximation we use relies on the smallness of the cone angle. This allows us to replace the Bessel function $J_0\left(\left|\vec{\rho}_{tip}' - \vec{\rho}_{tip}''\right| k\right)$ by its small argument series expansion. Numerically for typical values in our experiment we find that taking terms up to and including $k^4$ is sufficient. The final result for the MFM signal is given by:

$$\Delta f(z) = C - A\left[F_h(z + \lambda_{ab}) + F_{h+H}(z + \lambda_{ab} + H) - 2\overline{F}(z + H + \lambda_{ab})\right]. \quad [S3]$$

Here $C$ is an overall constant that sets the reference resonant frequency, $A \equiv f_0 \mu_0 m_0^2 \pi \tan^2 \theta / k_{cantilever}$ ( $f_0$ is the free-space cantilever frequency, $k_{cantilever}$ is the spring constant of the cantilever, $\mu_0$ is the permeability of the vacuum) and:

$$F_\beta(x) = \frac{1}{2}\left[\frac{1}{x} + \frac{\beta}{x^2} + \frac{\beta^2}{2x^3}\right] +$$
$$-\frac{3}{2}\tan^2\theta\left[\frac{1}{x} + \frac{\beta}{x^2} + \frac{3\beta^2}{4x^3} + \frac{\beta^3}{2x^4} + \frac{\beta^4}{4x^5}\right] + \quad [S4]$$
$$+\frac{3}{256}\tan^4\theta\left[\frac{256}{x} + \frac{256\beta}{x^2} + \frac{216\beta^2}{x^3} + \frac{176\beta^3}{x^4} + \frac{134\beta^4}{x^5} + \frac{90\beta^5}{x^6} + \frac{45\beta^6}{x^7}\right],$$

$$\overline{F}(x) = \frac{1}{2}\left[\frac{1}{x} + \frac{\frac{1}{2}H + h}{x^2} + \frac{h(H+h)}{2x^3}\right] +$$
$$-\frac{3}{2}\tan^2\theta\left[\frac{1}{x} + \frac{\frac{1}{2}H + h}{x^2} + \frac{3h^2 + 3hH + \frac{1}{2}H^2}{4x^3} + \frac{h^3 + \frac{3}{2}Hh^2 + \frac{3}{4}hH^2 + \frac{1}{8}H^3}{2x^4} + \right.$$
$$\left. + \frac{h^4 + 2Hh^3 + \frac{3}{2}h^2H^2 + \frac{1}{2}hH^3}{4x^5}\right] + \quad [S5]$$
$$+\frac{3}{8}\tan^4\theta\left[\frac{8}{x} + \frac{8(\frac{1}{2}H + h)}{x^2} + \frac{54h^2 + 54hH + 11H^2}{8x^3} + \frac{11(\frac{1}{2}H + h)^3}{2x^4} + \right.$$
$$\left. + \frac{134h^4 + 268h^3H + 198h^2H^2 + 64H^3h + 5H^4}{32x^5} + \frac{90(H+h)^5}{x^6} + \frac{45(H+h)^6}{x^7}\right].$$

## a. $H^*$ - the effective magnetic height of the tip

As mentioned in the main text we determine the parameter $H$ that appears in the model by finding the value for which our fit gives results with minimal sensitivity to the range of data we choose for the fit ($z_{span}$). Below we make a few observations and comments on the values we obtain and their systematics and speculate on the origin of the behavior.

The values we obtain ($H = H^*$) cover a range from roughly 20 $\mu m$, through 50 $\mu m$ to approximately 100 $\mu m$ with about 50% of the fits on the low end, 40% midrange and the rest on the high end. We find roughly the same value for $H^*$ for all data taken in a particular cool-down even when we measure more than one sample (recall that we measure each sample in multiple points). There were two separate cases of this kind: in one we measured two samples and in the other three. We also find that if we measure two samples from the same batch in different cool-downs we obtain the same result for the penetration depth even if the value of $H^*$ is different and even if we use a different tip. There was one such case where we did not change tips and one case where we did.

We speculate that the reason for this behavior of the parameter $H^*$ is that it is determined by the magnetic domain structure of the tip. As a result, once the tip is cold and we have magnetized it by applying a strong magnetic field (on the order of $0.5T$) the domain structure is fixed and we always get the same value of $H^*$. When we warm up, cool again and remagnetize – the domain structure can be different, and hence the value of $H^*$ can change.

## b. Test of the fit on YBCO

In order to validate our fitting procedure we implemented it for an optimally doped detwinned YBa$_2$Cu$_3$O$_{6.92}$ (YBCO) single crystal. In this sample our procedure gives $\lambda_{ab} \approx 160$ nm, smaller than any value we measured for the BaFe$_2$(AS$_{1-x}$P$_x$)$_2$ but larger by about 40nm than the value in the literature (120 nm [4]). This difference is not unreasonable considering that we did not cleave the YBCO sample and that it was stored for a long time at room temperature. Such samples are known to have a dead layer dozens of nanometers thick [4]. The formation of such a layer prevents the tip from reaching the actual superconducting surface, resulting in a fit value of $\lambda_{ab}$ larger by the thickness of the dead-layer.

## 2. Vortex-tip interaction versus the Meissner repulsion of the tip

Type II superconductors like the pnictides can contain superconductivity vortices. These vortices can interact with the magnetic tip of the MFM and change the resonance frequency. In this section we show that the tip-vortex interaction is much smaller than the Meissner response in our touchdowns. Because we are only interested in orders of magnitude we simplify the calculations by assuming $H \to \infty$ and by taking only the leading order in the cone angle $\theta$. We can account for both of these approximations by replacing $\vec{M}_{TCM}(x, y, z)$ in Eq. S1 by:

$$\vec{M}^{0}_{TCM}(x, y, z) = 2\pi m_0 \tan\theta (z+h)\, \delta(x)\delta(y)\, \Theta(z)\hat{z}. \qquad [S6]$$

In this model the tip looks like a needle in the $z$-direction, where the $2\pi(z+h)\tan\theta$ factor accounts for the increase in circumference of the tip with height, hence the increase of the number of magnetic moments per unit height.

### a. The Meissner repulsion

The result that the model in Eq. S6 gives is the leading order in $\tan\theta$ of Eqs. S4 & S5 for a tip with $H \to \infty$. Hence the Meissner response for this simplified model is:

$$\Delta f_M(z) = \frac{f_0 \pi (\tan\theta)^2 \mu_0 m_0^2}{2k}\left(\frac{1}{(z+\lambda_{ab})} + \frac{h}{(z+\lambda_{ab})^2} + \frac{h^2}{2(z+\lambda_{ab})^3}\right). \qquad [S7]$$

### b. The tip-vortex interaction

We use the Pearl approximation for the magnetic field of a vortex [5,6], which is valid when the vortex is far from the tip on a scale of $\lambda_{ab}$. the result of the approximation is that the magnetic field from a vortex is the field a monopole situated $\lambda_{ab}$ below the surface of the superconductor:

$$\vec{B}(r_v, z_v) = \frac{\Phi_0}{2\pi}\frac{r_v\hat{r} + (z_v + \lambda_{ab})\hat{z}}{\left[r_v^2 + (z_v + \lambda_{ab})^2\right]^{3/2}}, \qquad [S8]$$

where $\vec{r}_v = r_v\hat{r}$ is the in-plane position of the vortex, $z_v$ is the distance from the surface and $\Phi_0 = h/2e \approx 2.07\cdot 10^{-15}\, m^2 T$ is the flux quantum.

Using the Pearl model for a vortex (Eq. S8) and the model for the tip (Eq. S6) we can calculate an estimate of the MFM frequency shift due to vortex-tip interaction $\Delta f_v(r,z)$ as follows:

$$F_{v,z} = -\frac{\partial}{\partial z} U_{tip-vortex} = -\frac{\partial}{\partial z} \int \vec{M}_{tip} \cdot \vec{B}\, dv = -\Phi_0 m_0 \tan\theta \frac{\partial}{\partial z}\left( \int_0^\infty \frac{(z_v + z + \lambda_{ab})(z+h)}{\left[ r^2 + (z_v + z + \lambda_{ab})^2 \right]^{3/2}} dz_v \right) = $$
$$= \Phi_0 m_0 \tan\theta \frac{r^2 + (z+\lambda_{ab})(z+\lambda_{ab}+h)}{\left[ r^2 + (z+\lambda_{ab})^2 \right]^{3/2}},$$
[S9]

$$\Delta f_v(r,z) = -\frac{f_0}{2k} \frac{\partial F_{v,z}}{\partial z} = \frac{f_0 \Phi_0 m_0 \tan\theta}{2k} \frac{(z+\lambda_{ab})\left[ r^2 + (z+\lambda_{ab})^2 \right] + h\left[ 2(z+\lambda_{ab})^2 - r^2 \right]}{\left[ r^2 + (z+\lambda_{ab})^2 \right]^{5/2}}.$$
[S10]

### c. Meissner repulsion vs. interaction with a vortex

Typical values for our tips and samples are $\lambda_{ab} = 200\,\text{nm}$, $h = 50\,\text{nm}$, $\theta = \pi/9$ and $m_0 = 0.015\,\text{A}$. The value of $m_0$ which was calculated using the fit results for our data, also corresponds to a magnetic iron film coating approximately 10 nm thick. Using these values we plot in SFig. 1 the value of $\log_{10}(\Delta f_v/\Delta f_M)$ as a function of $r$ and $z$. The inset to SFig. 1 shows the value of the ratio (not the logarithm) at a cross-section of $r = 4$ μm. One can see that when the closest vortex is $r > 4\,\mu\text{m}$ away from the tip axis the effect is extremely minor. Our calculations show that for both effects to be of the same magnitude for $r > 4\,\mu\text{m}$ the tip magnetization has to be two orders smaller than it was in the experiment.

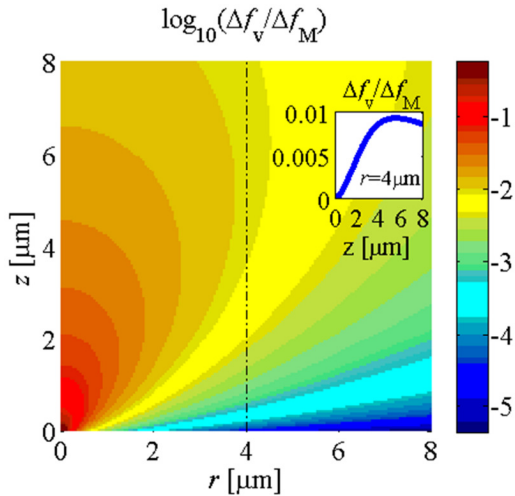

**Supplementary Figure 1:** The logarithm of the ratio of frequency shifts, $\log_{10}(\Delta f_v/\Delta f_M)$ ($\Delta f_v$ the frequency shift due to vortex-tip interaction, $\Delta f_M$ the shift due to the Meissner effect) for typical parameters ($\lambda_{ab} = 200\,\text{nm}$, $h = 50$ nm, $\theta = \pi/9$ and $m_0 = 0.015\,\text{A}$). **Inset:** $\Delta f_v/\Delta f_M$ as a function of height for $r = 4\,\mu\text{m}$ (marked in the main text by the black dot-dashed line).

## 3. Results for the penetration depth as a function of doping

Photographs of all the samples we used are shown in SFig. 2. These are all samples that were annealed in vacuum and cleaved just prior to mounting in the MFM. The samples with x=0.22 and x=0.26 were from two distinct batches, while samples D (x=0.29) and E (x=0.285) were two different pieces from the same crystal. Since MFM measurements are local, we measured each sample at 2 to 4 different positions. This provides a test of the uniformity of the samples. The EDS (Energy-Dispersive X-ray Spectroscopy) measurements with which we determined x were done at a beam energy of 20 kV and with a standard silicon reference sample. The EDS on each sample was performed at three different locations using a measurement area of about $50 \times 50\, \mu m^2$. We estimate the error for x for a particular sample from the variance of these EDS measurements.

Our results for the penetration depth are given in the supplementary table. The table also includes an extrapolation of the penetration depth to $T = 0$ K, denoted as $\tilde{\lambda}_0$, which we calculated using the temperature dependence data in Hashimoto et al. [7]. The results for the optimally doped YBCO sample (Sec. 1b) are also shown in the table.

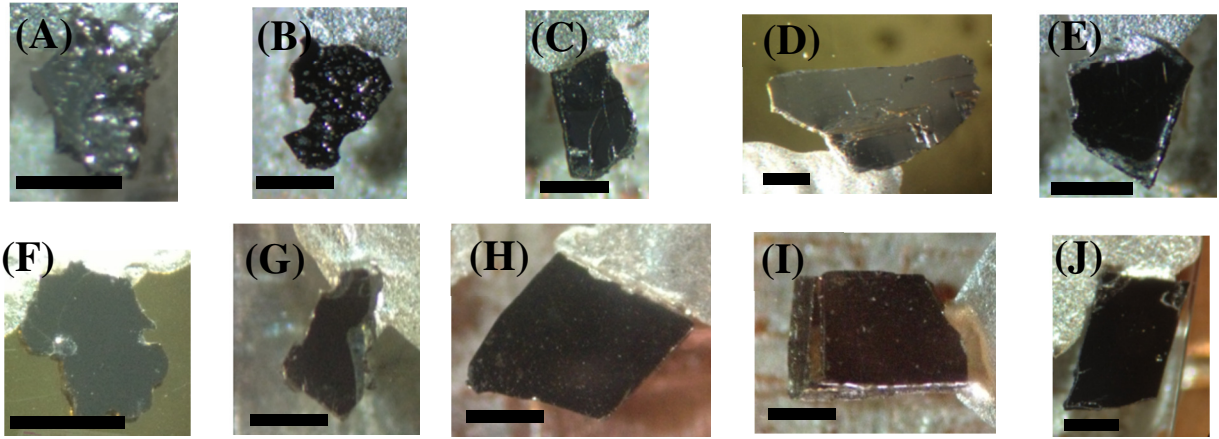

**Supplementary Figure 2:** Optical image of the BaFe$_2$(As$_{1-x}$P$_x$)$_2$ samples that we measured. The x value for each sample was **(A)** 0.550 **(B)** 0.48, **(C)** 0.30, **(D)** 0.29, **(E)** 0.285, **(F)** 0.33, **(G)** 0.26, **(H)** 0.26, **(I)** 0.22 and **(J)** 0.22. Samples I & J are from two different batches, while samples D & E and samples G & H are from the same batches. All the scale bars indicate 0.5mm.

**Supplementary Table:** Our estimates for $\lambda_{ab}$, the extrapolation to $T = 0K$ ($\tilde{\lambda}_0$) and the apparent $T_C$ for all samples and positions that appear in the main text. Errors in $\lambda_{ab}$ and x (the doping as measured by EDS) give a 70% confidence interval. The apparent $T_C$ is midway between the highest temperature at which we saw a Meissner effect and the lowest temperature where we did not see repulsion between the tip and the sample. Sample H (x=0.26) was measured uncleaved - hence its data was not used in the main text.

| sample | x | Site # | $\lambda_{ab}(T)$ [nm] | $T$ [K] | $\tilde{\lambda}_0$ [nm] | Apparent $T_C$ [K] |
|---|---|---|---|---|---|---|
| A | 0.55 ± 0.005 | 1 | 188 ± 10 | 4.6 | 165 | 11.0 ± 0.5 |
|   |   | 2 | 214 ± 10 | 4.6 | 190 | 11.0 ± 0.5 |
| B | 0.46 ± 0.01 | 1 | 201 ± 25 | 4.6 | 190 | 19.5 ± 0.25 |
|   |   | 2 | 214 ± 20 | 4.6 | 205 | 19.5 ± 0.25 |
| C | 0.30 ± 0.01 | 1 | 399 ± 15 | 4.5 | 385 | 31.0 ± 0.25 |
|   |   | 2 | 411 ± 10 | 4.5 | 400 | 30.5 ± 0.25 |
|   |   | 3 | 412 ± 15 | 4.5 | 400 | 30.5 ± 0.25 |
| D | 0.29 ± 0.01 | 1 | 300 ± 25 | 5 | 285 | 27.5 ± 0.5 |
|   |   | 2 | 268 ± 25 | 4.45 | 255 | 27.5 ± 0.5 |
|   |   | 3 | 261 ± 25 | 4.45 | 250 | 27.5 ± 0.5 |
| E | 0.285 ± 0.005 | 1 | 257 ± 25 | 4.5 | 245 | 27.0 ± 0.25 |
|   |   | 2 | 282 ± 25 | 4.45 | 270 | 27.0 ± 0.25 |
| F | 0.33 ± 0.01 | 1 | 383 ± 20 | 4.45 | 370 | 28.0 ± 0.5 |
|   |   | 2 | 372 ± 20 | 4.6 | 360 | 28.0 ± 0.5 |
|   |   | 3 | 372 ± 20 | 4.45 | 360 | 28.0 ± 0.5 |
| G | 0.26 ± 0.01 | 1 | 220 ± 20 | 4.45 | 200 | 22.0 ± 0.5 |
|   |   | 2 | 218 ± 20 | 4.65 | 200 | 21.5 ± 0.5 |
|   |   | 3 | 229 ± 25 | 4.6 | 210 | 21.5 ± 0.5 |
| I | 0.22 ± 0.02 | 1 | 764 ± 30 | 4.5 | 740 | 11.5 ± 0.5 |
|   |   | 2 | 446 ± 30 | 4.5 | 425 | 13.0 ± 0.5 |
| J | 0.22 ± 0.02 | 1 | 788 ± 30 | 4.55 | 765 | 10.5 ± 0.5 |
|   |   | 2 | 641 ± 35 | 4.6 | 620 | 10.5 ± 0.5 |
| YBCO | Near optimal doping | 1 | 157 ± 15 | 4.8 |   |   |
|   |   | 2 | 162 ± 15 | 4.8 |   |   |

## 4. Imaging

As mentioned in the main text, we used scanning to map the location of vortices with the goal of learning about the pinning landscape. Supplementary Figs. 3-5 show the results for different dopings. For large area scans, vortices might appear non-uniform and/or distorted if the scan plane was not exactly parallel to the surface.

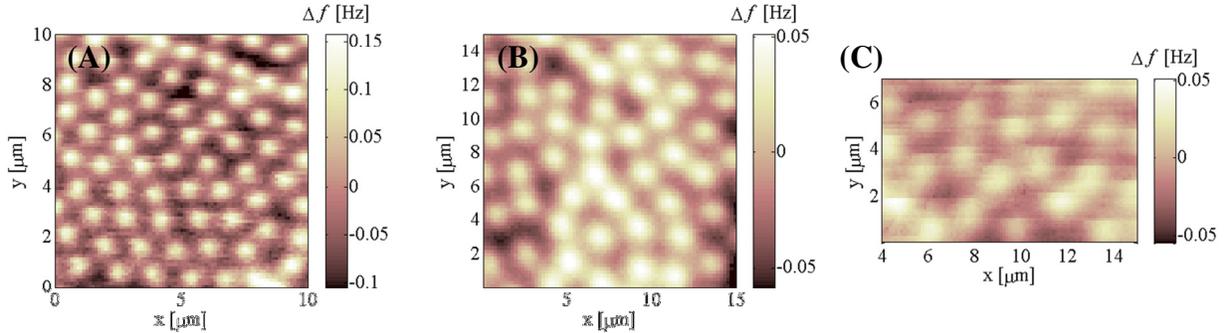

**Supplementary Figure 3:** Vortex decoration scans for overdoped samples. **(A)** Sample F (x=0.33) with $z$=600 nm, T=4.45 K, B=9 G (also shown in Fig. 4 in the main text); **(B)** Sample D (x=0.29) with $z$=900 nm, T=4.50 K, B=4 G; **(C)** Sample A (x=0.55) with $z$=850 nm, T=4.60 K, B=4 G.

As stated in the main text, in the very underdoped samples we could not see clear images of vortices. This can be seen in SFigs. 5A,B where the same area was scanned at different tip heights, $z$. Even with a large scan height the vortices appeared blurry (SFig. 5A), an indication of vortex motion. With the tip closer to the surface we obtained very streaky images indicating that vortices moved even more (SFig. 5B). A similar but larger area scan in SFig. 5C demonstrates the same kind of blurring at higher magnetic fields and in a different part of the same sample.

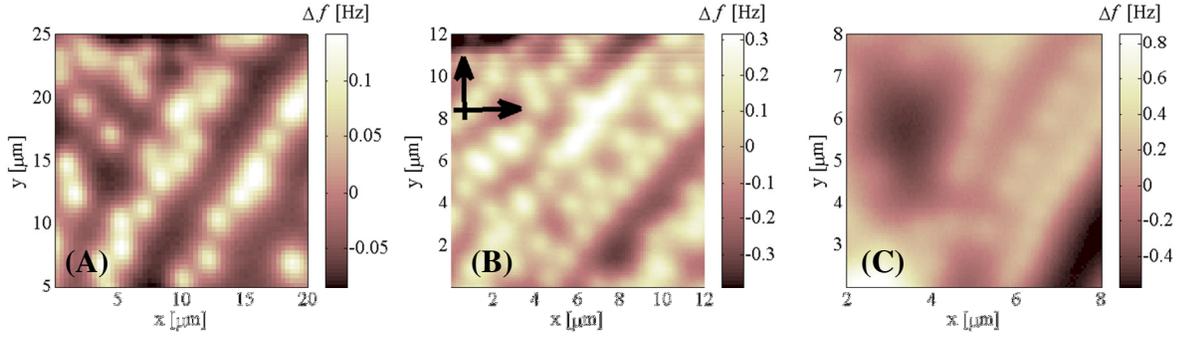

**Supplementary Figure 4:** Vortex decoration scans at x=0.26 showing lines of vortices. The scans shown here are from different samples and different areas. We show the direction of the *a,b* crystal axes where we determined them (black arrows). **(A)** Sample G – parallel lines are easy to discern (*z*=650 nm, T=4.45 K, B=5 G). **(B)** Different area in sample G also showing clear parallel lines (*z*=500 nm, T=4.65 K, B=10 G) (also shown in Fig. 4 in the main text). **(C)** Sample H showing straight lines of vortices (*z*=650 nm, T=4. 5 K, B=4 G). It appears that one of the lines is not parallel to the other two lines. This may be an artifact because the vortices are smeared. Unfortunately in this uncleaved sample we could not obtain clearer vortex scans. Because of this problem we did not use touchdowns from this sample in order to determine $\lambda_{ab}$.

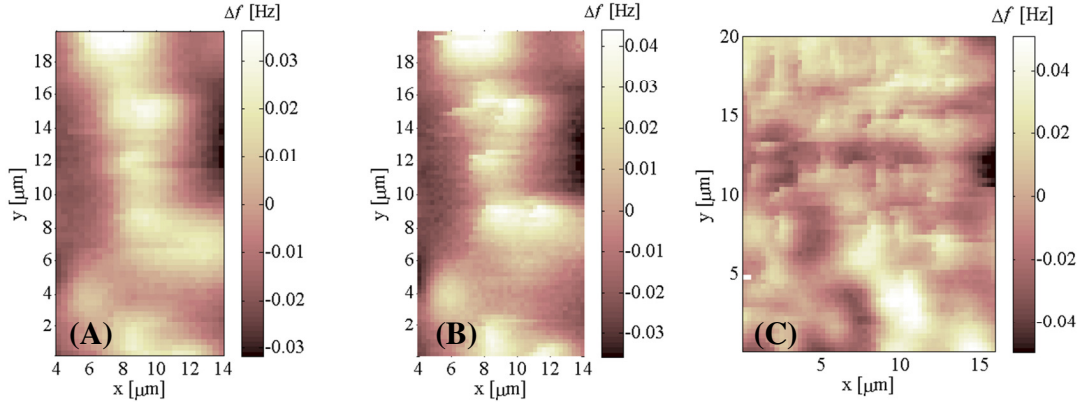

**Supplementary Figure 5:** Vortex decoration MFM scans for **(A)** Sample I (x=0.22) with *z*=100 nm, T=4.5 K, B=2 G; **(B)** Sample I (x=0.22) with *z*=850 nm, T=4.5 K, B=2 G; **(C)** Sample I (x=0.22) *z*=850 nm, T=4.5 K, B=5 G (also shown in Fig. 4 in the main text).

## 5. References for the supplementary material